# A decomposition of Fisher's information to inform sample size for developing or updating fair and precise clinical prediction models – Part 3: continuous outcomes

Rebecca Whittle[1,2*], Richard D Riley[1,2], Lucinda Archer[1,2,3], Gary S Collins[4], Amardeep Legha,[1,2] Kym IE Snell[1,2], Joie Ensor[1,2]

* Corresponding author:  r.l.whittle@bham.ac.uk

**Author details:**

[1] Department of Applied Health Sciences, School of Health Sciences, College of Medicine and Health, University of Birmingham, Birmingham, UK

[2] National Institute for Health and Care Research (NIHR) Birmingham Biomedical Research Centre, UK.

[3] Institute of Data and AI, University of Birmingham, UK

[4] Centre for Statistics in Medicine, Nuffield Department of Orthopaedics, Rheumatology and Musculoskeletal Sciences, University of Oxford, Oxford, OX3 7LD, UK

## Declarations

**Competing Interests**: RDR receives royalties for textbooks on prognosis research and IPD meta-analysis.

**Funding**: This paper presents independent research supported by an EPSRC grant for 'Artificial intelligence innovation to accelerate health research' (number: EP/Y018516/1), the MRC Better Methods Better Research panel (grant reference: MR/Z503873/1) and by the National Institute for Health and Care Research (NIHR) Birmingham Biomedical Research Centre at the University Hospitals Birmingham NHS Foundation Trust and the University of Birmingham. GSC is supported by Cancer Research UK (programme grant: C49297/A27294). RDR is partially supported by funding from the MRC-NIHR Methodology Research Programme [grant number: MR/T025085/1]. RDR and GSC are National Institute for Health and Care Research (NIHR) Senior Investigators. The views expressed are those of the author(s) and not necessarily those of the NHS, the NIHR or the Department of Health and Social Care.

**Author contributions:** RW derived the mathematical solutions, made the applied example, and wrote the initial draft of the manuscript. JE and RDR provided supervision and contributed to the conceptualization and development of the methodology, and critical review of the applied example and manuscript text. LA, GSC, AL and KIES provided critical feedback at multiple stages that led to the revision of the rationale, methods, applications, fairness checks, and text or figures in the manuscript. RW revised the manuscript after feedback from other authors. All authors agreed the final manuscript.

**Acknowledgements:** Not applicable

**Keywords:** Clinical prediction models, sample size, confidence intervals, instability, Fisher's information matrix, fairness, continuous outcomes

## Abstract

**Background**

Clinical prediction models enable healthcare professionals to estimate individual outcomes using patient characteristics. Current sample size guidelines for developing or updating models with continuous outcomes aim to minimise overfitting and ensure accurate estimation of population-level parameters, but do not explicitly address the precision of predictions. This is a critical limitation, as wide confidence intervals around predictions can undermine clinical utility and fairness, particularly if precision varies across subgroups.

**Methods**

We propose methodology for calculating the sample size required to ensure precise and fair predictions in models with continuous outcomes. Building on linear regression theory and the Fisher's unit information matrix, our approach calculates how sample size impacts the epistemic (model-based) uncertainty of predictions and allows researchers to either (i) evaluate whether an existing dataset is sufficiently large, or (ii) determine the sample size needed to target a particular confidence interval width around predictions. The method requires real or synthetic data representing the target population. To assess fairness, the approach can evaluate prediction precision across subgroups. Extensions to prediction intervals are included to additionally address aleatoric uncertainty.

**Results**

We demonstrate the methodology by examining the sample size required to develop or update a model for predicting Forced Expiratory Volume (FEV) using age, height, and sex. Existing guidance suggests a minimum sample size of 237 participants for this setting. We show this corresponds to an anticipated mean confidence interval width of 0.206 litres across all participants in the target population, and that widths may be considerably larger for some individuals. Our new approach calculates a minimum of 694 patients would be needed to ensure all anticipated interval widths were ≤0.3. Subgroup analysis showed comparable anticipated precision across sex subgroups. Sensitivity analysis assuming conditional independence among predictors yielded consistent results. For prediction intervals, the magnitude of the residual variance imposes a lower bound on interval width, even with very large samples.

## **Conclusions**

Our methodology provides a practical framework for examining required sample sizes when developing or updating prediction models with continuous outcomes, focusing on achieving precise and equitable predictions. It supports the development of more reliable and fair models, enhancing their clinical applicability and trustworthiness.

# Background

Clinical prediction models are vital tools in personalised medicine, supporting healthcare professionals in making informed decisions about diagnosis, prognosis, and treatment, by estimating an individual's likely outcome based on their characteristics and clinical information (i.e., 'predictors'). For instance, a model might predict a baby's expected birth weight at specific gestational ages using predictors such as maternal age, history of diabetes, and ethnicity (1). These predictions can help clinicians tailor interventions and monitor patients more effectively.

Accurate prediction models can help improve patient outcomes, optimise resource allocation, and enhance the overall quality of care. However, many existing models produce unreliable predictions, often due to flaws in study design and analysis (2–7). A common issue is the use of datasets that are too small for model development or updating, which can result in imprecise and poorly calibrated predictions (8,9). This, in turn, reduces clinical utility and may lead to flawed decision-making in practice (10). For this reason, the TRIPOD+AI reporting guideline specifically asks authors of prediction model studies to explain how the sample size was arrived at and justify it (11).

When developing models for predicting a continuous outcome, current guidance on sample size calculations suggests that the minimum sample size should aim to minimise model overfitting and ensure precise estimation of the overall mean value and the residual variance (12).

# Background

However, these guidelines do not account for the uncertainty in predictions, which can remain substantial due to model instability, even when the recommended minimum sample size is met (8,9,13). Large uncertainty results in wide confidence intervals around the predicted outcomes, reducing the model's reliability for clinical use. The acceptable level of uncertainty will depend on the specific context and outcome of interest.

To ensure that a model provides sufficiently precise predictions for decision making, sample size calculations are needed that specifically target the desired level of precision in outcome estimates. This is particularly important in clinical scenarios where decisions informed by predictions carry significant consequences.

A related consideration is the precision of predictions within relevant subgroups, especially those including underserved populations (e.g., ethnicity, socio-economic status) (8). Ensuring adequate precision across these groups is essential for fairness and equity in model performance.

In this article, we propose methodology for calculating sample sizes that target precise predictions when developing or updating models for continuous outcomes, building on our previous work for binary and time-to-event outcomes (14,15). After briefly summarising existing approaches, we extend them to derive closed-form solutions that can be applied either before data collection, or when an existing or pilot dataset is available. These solutions allow researchers to examine how the width of the confidence interval around a predicted outcome value varies with sample size, given the values of a specified set of predictors. The minimum required sample size can also be determined for a target confidence interval width.

To facilitate the approach, we describe methods for simulating datasets that resemble those that a researcher plans to collect, based on summary statistics from published studies. This enables practical implementation of our approach via our software module, *pmstabilityss*. We present applied examples to illustrate the methodology, followed by a concluding discussion. The solutions relate to minimising epistemic uncertainty (model-based error), but we extend to also consider prediction intervals (epistemic and aleatoric uncertainty).

## Existing sample size approach for developing a prediction model with a continuous outcome

We have previously derived calculations for the minimum sample size needed to predict continuous outcomes based on targeting the following criteria (12):

Criterion (i): a global shrinkage factor $\geq 0.9$

Criterion (ii): a small absolute difference in $R^2_{adj}$ and $R^2_{app}$

Criterion (iii): precise estimation of the residual standard deviation

Criterion (iv): precise estimation of the mean predicted outcome value (model intercept)

Further details of these sample size calculations are available in previous papers (12,16). The approach is implemented in the *pmsampsize* package (available in Stata, R and Python) (17,18), with the user needing to specify the anticipated $R^2$ of the model, the average outcome value, and the standard deviation of outcome values in the population of interest.

For example, in the applied example we consider the development of a linear regression model to estimate Forced Expiratory Volume (FEV) in children based on three predictor parameters: age, sex and height (assuming linear associations for age and height). If we assume that an existing prediction model in the same field has an $R^2$ of 0.77, and that FEV values in the target population have a mean of 2.64 litres and standard deviation of 0.87 litres, then *pmsampsize* calculates that at least 237 participants are needed to meet criteria (i) to (iv).

# New sample size approach to target precise estimates of continuous outcomes

We now introduce a new method to help researchers assess how the sample size for model development may influence the precision and fairness of outcome estimates. The approach is grounded in linear regression and maximum likelihood theory and is also applicable when extending or updating an existing model, especially when existing or pilot data are available. This is because it requires researchers to input information on anticipated case-mix distributions and model performance. The method provides an initial target sample size that can inform, for example, grant applications or study planning. We emphasise that once data collection and analysis is underway, internal validation remains essential to assess model stability and performance, i.e., by using learning curves and stability plots (8,19).

## Step 1: Identify the predictors expected to be included in the model

In addition to sample size and model performance (see later), the uncertainty of a prediction for a continuous outcome depends on the predictors included in the model. Therefore, to estimate prediction uncertainty, it is essential to identify the predictors most likely to be included in the model being developed; referred to here as the *core predictors*. The set of core predictors are typically those with well-established predictive value in the relevant clinical context but may also include emerging predictors of interest.

Core predictors can be identified through previously published models, systematic reviews of prognostic factors (20), or in consultation with clinical experts. For example, when predicting systolic blood pressure (SBP), core predictors might include age, sex, BMI, ethnicity, and diabetes status (21–24) .

While additional predictors may be considered during model development, potentially increasing the uncertainty of predictions, the core predictors and their value combinations provide a foundational basis for assessing sample size and the precision of predictions.

## Step 2: Identify or create a relevant dataset

The joint distribution of the predictors included in the final prediction model affects the standard errors of the linear regression model's parameter estimates, thereby influencing the width of the confidence intervals around predictions. Therefore, in Step 2, the joint distribution of the core predictors identified in Step 1 must be specified. The method for doing this depends on the information available at the time of sample size calculation:

*Scenario 1: Existing dataset or pilot data are available to the model development team*

This is a common situation when a team is planning to develop or update a prediction model. It represents the simplest case for calculating sample sizes using the proposed method, as the dataset containing the predictors and their joint distribution is already available to be used for the sample size calculation. In this scenario, the user can proceed directly to Step 3.

*Scenario 2: No existing data are available*

If individual-level data are not currently available to the model development team, data can be simulated to approximate the joint distribution of the core predictors.

If a previously published article reports results from both univariable and multivariable linear regression models that include the core predictors of interest for a population similar to the target population, this information can be used to estimate the joint distribution of the core predictors. Specifically, the covariance between each core predictor, $x$, and the outcome, $y$, can be calculated using the reported regression coefficient, $\beta_{uni}$, from a univariable regression of the outcome on the predictor, using the following formula:

$$cov(x, y) = \beta_{uni} \times var(x). \qquad (1)$$

This assumes that the variance of the predictor is either reported or can be calculated (e.g., from a reported standard deviation), and that all predictors are continuous. If binary predictors are included (including dummy variables for categorical predictors), their variance can be calculated as $p(1-p)$, where $p$ is the proportion of individuals with the predictor value equal to 1.

Once the covariances between each predictor and the outcome are obtained, they can be combined with the vector of regression coefficients from a multivariable model, $\boldsymbol{\beta_{multi}}$, to estimate the covariances between the predictors using:

$$Cov(\boldsymbol{X}) = \boldsymbol{\beta_{multi}^{-1}}\big(Cov(X_i, y)\big)'. \qquad (2)$$

The variance-covariance matrix in Equation (2) can be solved using simultaneous equations and can then be used to simulate a dataset reflecting the joint distribution of the core predictors in the reference population.

If a relevant dataset exists but is not yet available (e.g., access is conditional on funding approval), the data holders may be able to provide a synthetic dataset that mimics the joint distribution of the predictors. For example, the Clinical Practice Research Datalink (CPRD) has previously provided synthetic datasets to support protocol development. Alternatively, data holders may provide the variance-covariance matrix of the predictors, allowing the research team to simulate a dataset that reflects the expected joint distribution, without needing to use approximations.

If it has not been possible to obtain information on the joint distribution of the predictors, either because a new dataset needs to be collected or because data holders were unable to provide the necessary information, researchers may still simulate a data using available summary statistics (e.g., means, standard deviations, or proportions for categorical variables). These can often be extracted from published tables of baseline characteristics. In this case, the predictors are assumed to be conditionally independent, meaning no correlations are specified between them. The impact of this assumption might be explored through sensitivity analyses, for example by simulating datasets under different assumed correlation structures. We consider this in an example later.

### Step 3: Derive the Fisher's unit information matrix

Unlike in logistic regression and survival models (14,15), the precision of a prediction from a linear regression model is not conditional on the actual predicted value itself. However, it still depends on the model's overall performance (e.g., explained variation), the sample size, and the individual's predictor values.

To disentangle the effect of sample size from these other elements, Step 3 uses a decomposition of the variance of the regression coefficients (for the core predictors), $var(\widehat{\boldsymbol{\beta}})$ (the inverse of the Fisher's information matrix), into Fisher's *unit* information matrix, $\mathbf{I}(\boldsymbol{\beta})$, and the sample size, $n$. Specifically, the variance of the estimated regression coefficients $(\widehat{\boldsymbol{\beta}})$ is given by:

$$var(\widehat{\boldsymbol{\beta}}) = \frac{1}{n}\mathbf{I}(\boldsymbol{\beta})^{-1} \qquad (3)$$

where:

$$\mathbf{I}(\boldsymbol{\beta}) = \frac{1}{\sigma^2} E(\boldsymbol{X}^T\boldsymbol{X}). \qquad (4)$$

Here, $X$ is the design matrix for the core predictors, with each row representing an individual's data $x_i = (1, x_{1i}, x_{2i}, \ldots, x_{pi})$ for $p$ core predictors, and $\sigma^2$ is the residual (unexplained) variance of the outcome.

To use Equation (3) for sample size calculation, we must specify the anticipated value of $\mathbf{I}(\boldsymbol{\beta})$. This depends on the expected value of $X^TX$ (derived in Step 2) and the residual variance $\sigma^2$, which must be specified by the researcher. Alternatively, the anticipated $R^2$ of the model can be specified, which is related to the residual variance through the following expression:

$$\sigma^2 = \frac{(1-R^2)\sum(y_i - \bar{y})^2}{N}, \qquad (5)$$

where $y_i$ is the value of the outcome for individual $i$, $\bar{y}$ is the mean value of the outcome, and $N$ is the size of the dataset identified or created in Step 2.

Substituting this into Equation (4), the Fisher's unit information matrix can be expressed as:

$$\mathbf{I}(\boldsymbol{\beta}) = \frac{X^TX}{(1-R^2)\sum(y_i - \bar{y})^2}. \qquad (6)$$

Thus, to compute the Fisher's unit information matrix, we need:

- The $X^TX$ matrix derived from the data in Step 2

and, either:

- an assumed value of the residual variance, $\sigma^2$

or:

- an assumed value for $R^2$, and
- the sum of squares of the outcome, $\sum(y_i - \bar{y})^2$.

If a dataset is not available, the outcome variable $y$ can be simulated using a known or assumed mean and standard deviation, to estimate $\sum(y_i - \bar{y})^2$. Importantly, the outcome in the simulated data does not need to be conditionally dependent on the predictors, as the Fisher's unit information matrix is based only on the predictor distribution and residual variance, not on the predictor-outcome relationship.

## Step 4: Examine the impact of sample size on the precision of outcome estimates

The final step involves assessing how the size of the dataset intended for model development or updating may affect the precision of outcome predictions, as reflected in the width of anticipated 95% confidence intervals for expected values. This is particularly relevant when working with an existing dataset, either to determine whether it is sufficiently large, or to use it as pilot data to inform prospective data collection and identify appropriate target sample sizes.

*Option A: Calculate the anticipated confidence interval widths of predictions for a given sample size*

This option applies when a dataset already exists for model development. The variance of the predicted (expected) outcome value ($E(y_{new}) = \hat{y}_{new}$) for new individuals with the same set of predictor values is given by:

$$var(\hat{y}_{new}) = var(\boldsymbol{x}_{new}\widehat{\boldsymbol{\beta}}) = \boldsymbol{x}_{new}var(\widehat{\boldsymbol{\beta}})\boldsymbol{x}'_{new}, \qquad (7)$$

where $\boldsymbol{x}_{new} = (1, x_{1,new}, x_{2,new}, \ldots, x_{p,new})$ are the predictor values for the new individuals.

Substituting Equation (3) from Step 3, we get:

$$var(\hat{y}_{new}) = \frac{1}{n}\boldsymbol{x}_{new}\mathbf{I}(\boldsymbol{\beta})^{-1}\boldsymbol{x}'_{new}. \qquad (8)$$

If we had an estimate of the individual's predicted outcome, we could derive an anticipated confidence interval around this estimate using:

$$\left[\hat{y}_{new} \pm \left(t_{a/2,n-p}\sqrt{var(\hat{y}_{new})}\right)\right] = [\hat{y}_{new_L}, \hat{y}_{new_U}], \qquad (9)$$

where $t_{a/2,n-p}$ is the critical value from the t-distribution with $n - p$ degrees of freedom, $n$ is the sample size, $p$ is the number of parameters, and L and U denote the lower and upper bounds, respectively.

For large sample sizes (typically n>100, which we would expect when using *pmsampsize* as a starting point), the $t$-distribution can be approximated by the standard normal distribution, where $z_{\alpha/2}$ is the critical value from the standard normal distribution.

Although the predicted value $\hat{y}_{new}$ is unknown at the sample size calculation stage, we can use Equation (8) to compute the width of the confidence interval (CI) as:

$$\text{CI width} = 2 \times t_{a/2, n-p-1} \sqrt{var(\hat{y}_{new})}$$

$$= 2 \times t_{a/2, n-p-1} \sqrt{\frac{1}{n} (x_{new} \mathbf{I}(\boldsymbol{\beta})^{-1} x'_{new})}. \tag{6}$$

This allows the user to evaluate the anticipated confidence interval widths for individuals in the existing or simulated dataset (from Step 2), using the Fisher's unit information matrix **I(β)** from Step 3 and a specified sample size $n$. These widths can then be assessed for clinical acceptability, either overall, or within specific subgroups.

*Option B: Calculate a target sample size for new data collection*

When planning a new study, researchers may wish to determine the sample size required to achieve a desired level of precision in outcome predictions. Rearranging Equation (8), the required sample size to achieve a target prediction variance is:

$$n = \frac{1}{var(\hat{y}_{new})} x_{new} \mathbf{I}(\boldsymbol{\beta})^{-1} x'_{new}. \tag{7}$$

Alternatively, this can be expressed in terms of a target confidence interval width, assuming a normal distribution:

$$n = \frac{x_{new} \mathbf{I}(\boldsymbol{\beta})^{-1} x'_{new}}{\left(\frac{\text{CI width}}{2 \times z_{\alpha/2}}\right)^2}. \tag{12}$$

This calculation can be applied to each individual in the dataset from Step 2 to determine the required sample size for their specific predictor profile $x_{new}$, given a maximum acceptable confidence interval width for the outcome of interest.

## Applied Example

To illustrate the proposed methodology, we consider the development of a prediction model for estimating Forced Expiratory Volume (FEV) in children and adolescents. The goal is to determine the sample size required to ensure precise predictions of FEV from a linear regression model, with age, sex, and height as predictors. Moreover, an existing dataset containing lung function measurements from 654 individuals is freely available (25), and so the calculations aim to examine the anticipated precision of predictions from this dataset (in advance of analysis), to help decide (e.g. in a grant funding application) whether it is fit for purpose.

We explore two practical scenarios: one in which researchers have access to the existing dataset and can use it directly to inform the sample size calculations (Scenario 1), and another in which access to the raw data is not available, but relevant summary statistics can be obtained from published sources or data custodians (Scenario 2). These scenarios reflect common situations in clinical research and allow us to demonstrate how the proposed methodology can be flexibly applied depending on the level of data access. In both cases, our aim is to determine the minimum sample size required to achieve a target level of precision in predictions, as quantified by the average width of the confidence intervals, and to ascertain whether the existing dataset of 654 participants is likely appropriate. We now walk through the four steps outlined above.

## Step 1: Identify the predictors expected to be included in the model

For this example, we assume that age, height, and sex are the core predictors of interest for predicting FEV, and that they are linearly associated with the outcome. These variables are well-established in the literature as key predictors of lung function in children (26).

## Step 2: Identify or create a relevant dataset

*Scenario 1: Existing dataset or pilot data are available to the model development team*

We use the existing dataset of 654 children aged 3-19 years, collected in East Boston, United States, during the 1970's (25). This dataset includes information on FEV (in litres) and our three core predictors (age, height, and sex). It is openly available at https://hbiostat.org/data/. Since the joint distribution of the three core predictors can be directly observed in this dataset, no simulation is required, and we can proceed to Step 3.

*Scenario 2: No existing data are available*

Now suppose we do not have access to the individual-level data. Instead, we assume that the data owners are not willing to share their data until grant funding is obtained, but do provide the summary statistics and regression coefficients presented in Tables 1 and 2.

Table 1: Summary information for core predictors and the outcome available from previous publication

| Characteristic | N=654 |
|---|---|
| Age (years); *Mean (SD)* | 9.9 (2.95) |
| Height (inches); *Mean (SD)* | 61.1 (5.7) |
| Sex (male); *N (%)* | 336 (51.38%) |
| FEV (litres); *Mean (SD)* | 2.64 (0.87) |

Table 2: Model coefficients for the core predictors from a previous publication

|  | Univariable Coefficient (95% CI) | Multivariable Coefficient (95% CI) |
|---|---|---|
| Age (years) | 0.222 (0.207, 0.237) | 0.061 (0.044, 0.079) |
| Height (inches) | 0.132 (0.126, 0.138) | 0.105 (0.095, 0.114) |
| Sex (male) | 0.361 (0.231, 0.492) | 0.161 (0.096, 0.226) |

Using the standard deviations (to calculate the variance) from Table 1 and the univariable coefficients from Table 2, we calculate the covariances between each predictor and the outcome using Equation (1). We then use these covariances, along with the multivariable coefficients, to solve the system of equations derived from Equation (2) to estimate the variance-covariance matrix of the predictors:

$$Cov(\boldsymbol{X}) = \begin{pmatrix} var(x_{age}) & cov(x_{age}, x_{height}) & cov(x_{age}, x_{sex}) \\ cov(x_{age}, x_{height}) & var(x_{height}) & cov(x_{height}, x_{sex}) \\ cov(x_{age}, x_{sex}) & cov(x_{sex}, x_{height}) & var(x_{sex}) \end{pmatrix}.$$

This results in the following system of equations:

$$\beta_{age} var(x_{age}) + \beta_{height} cov(x_{age}, x_{height}) + \beta_{sex} cov(x_{age}, x_{sex}) = cov(x_{age}, y)$$

$$\beta_{age} cov(x_{age}, x_{height}) + \beta_{height} var(x_{height}) + \beta_{sex} cov(x_{sex}, x_{height}) = cov(x_{height}, y)$$

$$\beta_{age} cov(x_{age}, x_{sex}) + \beta_{height} cov(x_{height}, x_{sex}) + \beta_{sex} var(x_{sex}) = cov(x_{sex}, y)$$

These equations can be solved using software such as R (using *solve()*) or Python (using *sym.Eq*). Once the variance-covariance matrix is obtained, we assume multivariate normality to simulate a dataset using the *mvrnorm()* function in R, generating n=1,000,000 observations (R code to replicate the example can be found at https://github.com/RebeccaWhittle).

Since sex is a binary variable, we convert the simulated continuous values into binary by sorting and assigning the top 48.62% as female (sex = 0) and the remainder as male (sex = 1), based on the observed proportion in Table 1.

We also simulate the outcome variable (FEV) from a normal distribution with the reported mean and standard deviation. Importantly, this outcome does not need to be conditionally dependent on the predictors for the purpose of estimating the Fisher's unit information matrix in Step 3.

### Step 3: Derive the Fisher's unit information matrix

With a (synthetic) dataset now available, either the original dataset of 654 children or a large dataset simulated in Step 2, we proceed to compute the Fisher's unit information matrix, $\mathbf{I}(\boldsymbol{\beta})$, using Equation (6).

For illustrative purposes, we assume an anticipated $R^2$ value of 0.77, chosen to reflect a plausible level of explained variance for a model including age, height, and sex in this clinical context. This value was based on previous prediction models for FEV and previous analyses using subsets of the same data, which is akin to when using a pilot study for new data collection.

The matrix $\boldsymbol{X}^T\boldsymbol{X}$ is calculated from the design matrix of the core predictors (age, height, sex), and the sum of the squares of the outcome $\sum(y_i - \bar{y})^2$ is computed using the observed or simulated FEV values.

# Step 4: Examine the impact of sample size on the precision of outcome estimates

Finally, both options A and B of Step 4 were applied to the two scenarios of Step 2 to examine and determine the sample size required for appropriately narrow confidence intervals.

*Option A: Calculate the anticipated confidence interval widths of predictions for a given sample size*

Option A can be used to calculate the anticipated confidence intervals for predicted outcomes in either the real or simulated dataset from Step 2, conditional on a user-specified sample size. This requires the Fisher's unit information matrix derived in Step 3, which is then applied in Equation (10) alongside each individual's predictor values and the specified sample size to compute the width of the anticipated confidence intervals.

In Scenario 1 (when using the existing dataset), we first calculated the anticipated confidence interval widths for the fixed sample size of n=654, corresponding to the number of observations in the available dataset. This yielded a mean interval width of 0.124, with a range of 0.089 to 0.309. In Scenario 2 (using the simulated dataset), the same sample size of n=654 produced an identical mean width of 0.124, though with a slightly wider range of interval widths (0.089 to 0.352).

We also evaluated the confidence interval widths for a smaller sample size of n=237, which corresponds to the minimum required sample size calculated using the Riley et al. criteria described above (12,17,18). As expected, the average confidence intervals were wider, by approximately 0.08 units compared to the larger dataset (Scenario 1: 0.206 (0.148 to 0.513); Scenario 2: 0.207 (0.148 to 0.583)). These results are visualised in Figure 1, which compares the distribution of confidence interval widths at both sample sizes. While the increase is consistent with theoretical expectations, the absolute difference in this example is small, reflecting the relatively narrow confidence intervals in both sample sizes.

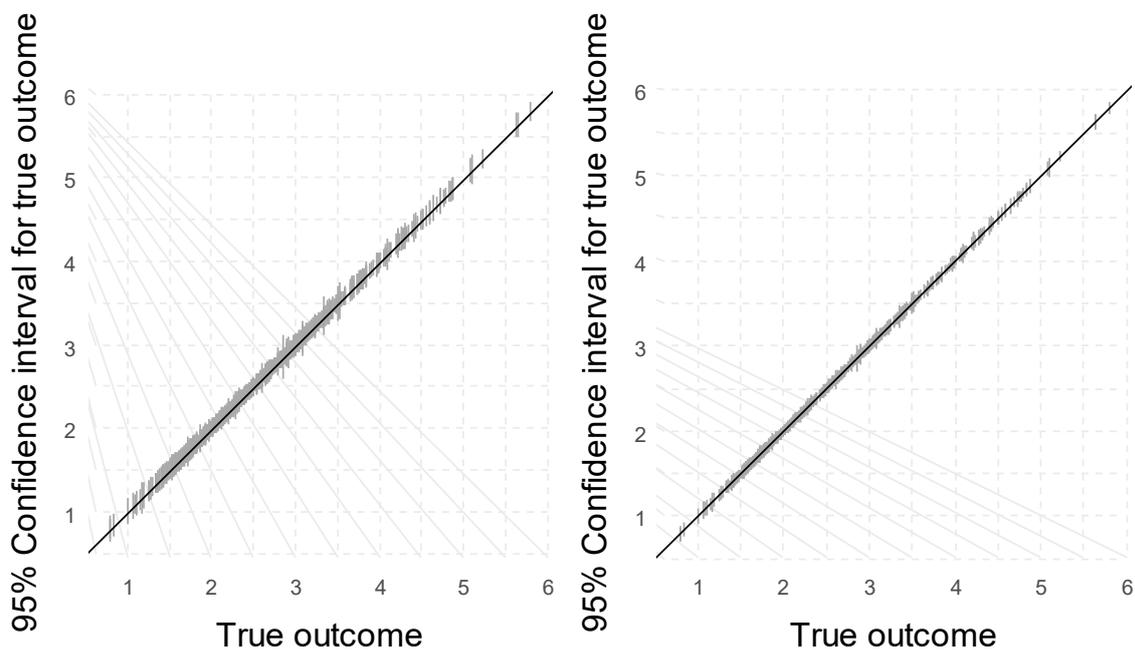

*Figure 1: Estimated confidence interval widths for predicted values of FEV in the existing dataset when the sample size is A) N=237 and B) N=654*

### Option B: Calculate a target sample size for new data collection

We now apply Option B to calculate the minimum sample size required to ensure that *all* predictions have confidence intervals no wider than 0.3 units, which was considered a reasonably acceptable width for this particular clinical scenario. This is achieved by applying Equation (12) to each individual in the dataset.

Using the existing dataset (Scenario 1), the calculations indicate that a minimum sample size of 694 participants would be required to ensure that all predicted outcomes have confidence interval widths ≤ 0.3. When using the simulated dataset (Scenario 2), the estimated minimum sample size increases slightly to 733 participants, reflecting minor differences in the simulated and real data.

To further illustrate the relationship between sample size and prediction precision, we plotted the minimum sample size required to ensure that all anticipated confidence intervals were at least as narrow as the target width specified on the x-axis (see Figure 2). The plot shows that at n=237, the minimum recommended by *pmsampsize*, all anticipated confidence interval widths are below 0.51 units, but many are not below 0.3. This highlights the added value of our approach in tailoring sample size calculations to target the precision of predictions.

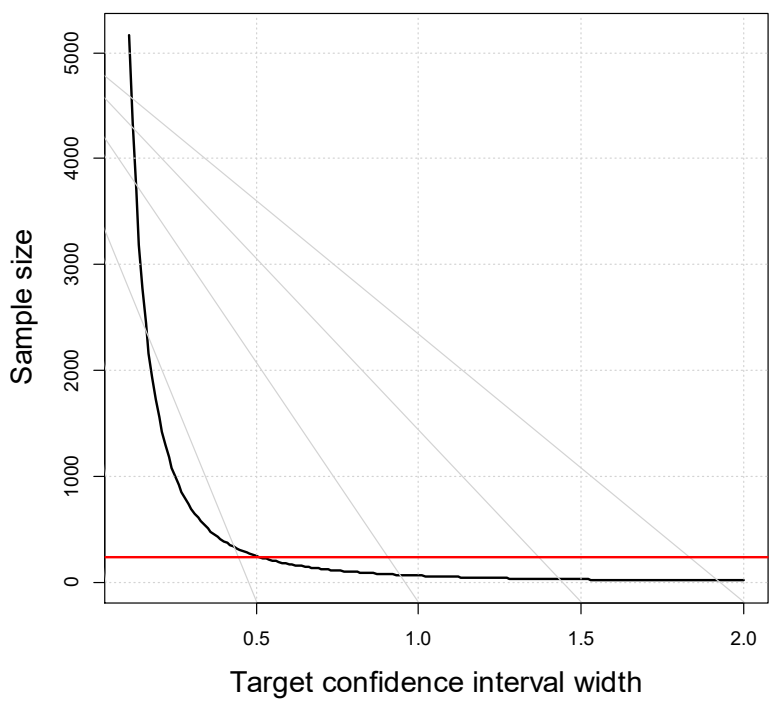

*Figure 2: Minimum sample size required to ensure confidence interval widths for all predictions are below a specified threshold in the FEV example. The red line indicates the sample size of 237 (as calculated using pmsampsize).*

# Extensions

## Evaluating the confidence interval widths within subgroups

To demonstrate how prediction precision may vary across subgroups, we conducted a stratified analysis by sex. In this context, we focus on one specific aspect of fairness: the consistency of prediction precision across groups. While fairness in clinical prediction models encompasses broader concerns, including biases in data and unequal outcomes, our analysis specifically examines whether one subgroup consistently experiences wider confidence intervals. Such differences in precision may indicate that the model is less reliable for that group, potentially influencing clinical decision making and contributing to inequitable access to healthcare resources.

Using the existing dataset (n=654), we calculated the confidence interval widths separately for male and female participants. This allowed us to evaluate whether the sample size was sufficient to ensure comparable levels of precision across both groups. The mean confidence interval width for females was 0.122 (range: 0.091 to 0.309), and for males, it was 0.125 (range: 0.089 to 0.236). These findings suggest that the model will be expected to achieve a comparable level of precision in predictions across sexes, with only minimal differences in confidence interval widths, which is reassuring in terms of fairness of predictions for males and females.

## Assuming predictors are conditionally independent

One of the most significant practical challenges in applying the proposed methodology is the need to specify the joint distribution of the core predictors. This information is essential for calculating the Fisher's unit information matrix and, in turn, the uncertainty associated with outcome predictions. However, in many real-world scenarios, researchers may only have access to marginal summary statistics (e.g., means, standard deviations, or proportions), without information on the correlations between predictors.

To assess the potential impact of this limitation, we conducted a sensitivity analysis in which we simulated a dataset under the assumption that the core predictors (age, height, and sex) were conditionally independent. This allowed us to evaluate how much the assumption of independence might affect the resulting confidence interval widths.

For a sample size of 654, under the assumption of independent predictors, the confidence intervals had a mean width of 0.124, with a range from 0.088 to 0.351. For a smaller sample size of 237, the mean width increased to 0.206, with a range from 0.148 to 0.583.

These results were highly consistent with those obtained using the true joint distribution of the predictors, suggesting that assuming conditional independence may be a reasonable approximation when full joint distribution information is unavailable. This finding provides reassurance that the methodology can still be applied in settings where only limited information is available, though we recommend that researchers conduct similar sensitivity analyses to assess robustness in their specific context.

## Prediction intervals

So far, our calculations have considered widths of confidence intervals about the expected (mean) value of a prediction given a new individual's set of predictor values. As the focus is on the uncertainty of the mean value, the confidence interval width only acknowledges model-based error, also known as epistemic uncertainty. However, researchers may also be interested in the *total* uncertainty of a prediction for a new individual, which should acknowledge not only epistemic uncertainty but also aleatoric uncertainty. The latter reflects irreducible variability in the outcome that cannot be explained by the set of (candidate) predictors included in the model, and leads to a prediction interval which will be wider than a confidence interval.

Formally, a prediction interval provides a range within which we expect an individual's future observed outcome value to fall, accounting for both model uncertainty and residual error. Unlike confidence intervals around *expected* (mean) outcome values, prediction intervals for individual outcome values do not tend to zero even with infinite sample size, because the residual variance remains.

A prediction interval for a new individual's actual outcome value ($\hat{y}_{new}$) is approximately calculated as

$$\hat{y}_{new} \pm \left(t_{a/2, n-p-1}\sqrt{var(\hat{y}_{new}) + \sigma^2}\right), \tag{13}$$

where $var(\hat{y}_{new})$ is calculated as in Equation (8), $\sigma^2$ is the residual variance, estimated using Equation (5), and $t_{a/2,n-p-1}$ is the critical value from the t-distribution with $n-p-1$ degrees of freedom where $n$ is the sample size and $p$ is the number of predictor parameters (excluding the intercept). For large sample sizes (typically n>100, which we would expect when using *pmsampsize* as a starting point), the *t*-distribution can be approximated by the standard normal distribution, yielding:

$$\hat{y}_{new} \pm \left(z_{\alpha/2}\sqrt{var(\hat{y}_{new}) + \sigma^2}\right). \quad (14)$$

This expression can be rearranged to estimate the required sample size to achieve a target prediction interval (PI) width:

$$n = \frac{x_{new}\mathbf{I}(\boldsymbol{\beta})^{-1}x'_{new}}{\left(\frac{\text{PI width}}{2 \times z_{\alpha/2}}\right)^2 - \sigma^2}. \quad (15)$$

where $z_{\alpha/2}$ is the critical value from the standard normal distribution.

However, this formula has a critical constraint: the denominator must be positive. That is,

$$\left(\frac{\text{PI width}}{2 \times z_{\alpha/2}}\right)^2 > \sigma^2.$$

This highlights a key property of prediction intervals: the minimum achievable prediction interval width is bounded by the residual variance. In other words, no matter how large the sample size is, even approaching infinity, the aleatoric uncertainty cannot be eliminated. As a result, the prediction interval cannot be made arbitrarily narrow. This underscores the importance of considering the magnitude of residual variance when planning the development of a prediction model.

To illustrate, we return to the applied example. Using the existing dataset (Scenario 1), the residual variance was calculated to be 0.173, and so across all individuals the mean 95% prediction interval width was estimated as 1.634, with a range from 1.631 to 1.658 for a sample size of n=654. For a smaller sample size of n=237, the mean width was 1.642, ranging from 1.635 to 1.708. However, even with n=100,000,000 (i.e., eradicating the epistemic uncertainty), the mean prediction interval width is only marginally reduced to 1.629. This result highlights that when aleatoric uncertainty (i.e. residual variance) is present, increasing the sample size can have only a limited impact on the width of prediction intervals as the uncertainty due to residual variance will always remain. For this reason, generally we recommend to focus sample size calculations on the target confidence interval width.

## Discussion

The reliability of a prediction model is intrinsically linked to the sample size used in its development (10). While larger sample sizes generally lead to more stable and accurate models, datasets that meet current sample size recommendations, such as those ensuring small optimism in predictor effect estimates, minimal difference between apparent and adjusted $R^2$, and precise estimation of the residual standard deviation and model intercept (as outlined by Riley at. (12)), can still yield substantial uncertainty in predictions (8,9).

In this article, we have extended existing guidance by proposing calculations to determine the minimum sample size required to target precise predictions for models with continuous outcomes. Our approach builds on previous work for binary (14) and time-to-event outcomes (15), and introduces a framework for calculating sample sizes based on target width of the confidence or prediction intervals for individuals in the target population. We provide methods applicable both when individual-level data are available and when only summary statistics can be accessed. As our method assumes unbiased and well-calibrated predictions, we recommend using it in conjunction with existing tools such as *pmsampsize* to ensure that criteria to minimise overfitting and that aim for well-calibrated predictions at the population level are met (12).

In addition to model development, the proposed methodology is also applicable to model updating. Researchers can use information from a previously published model and an existing dataset to assess the expected uncertainty in updated predictions, or to determine the sample size required to achieve a desired level of precision in the updated model.

Researchers can use this framework in two ways: (i) to evaluate whether an existing dataset is sufficiently large to yield clinically acceptable prediction precision, or (ii) to calculate the required sample size for new data collection based on a target level of prediction precision. Importantly, our approach also supports fairness considerations by enabling researchers to assess and target prediction precision within specific subgroups, such as those defined by age, sex, ethnicity, or socioeconomic status, thereby working towards promoting parity in model performance. Ensuring fairness in clinical prediction models is increasingly recognised as a critical component of responsible model development (11). A model that performs well on average may still yield systematically less precise or less accurate predictions for certain subgroups, potentially exacerbating existing health disparities. By applying the sample size calculations separately within each subgroup, researchers can evaluate whether the available data are sufficient to ensure acceptable confidence interval widths across all relevant populations. This is particularly important when subgroups are underrepresented in the dataset or when predictor-outcome relationships differ meaningfully between groups. However, assessing the consistency of confidence intervals across subgroups is only one aspect of fairness, and even with this approach, it does not guarantee that the resulting models will be fully fair or equitable.

Our method is grounded in maximum likelihood theory for linear regression and uses the Fisher's unit information matrix to quantify epistemic (reducible model-based) uncertainty (27) via confidence intervals around the mean (expected) predicted value for a new individual. The method focuses on quantifying and minimising this epistemic uncertainty, which arises from estimating model parameters using finite data. This is captured through the Fisher's unit information matrix and reflects the variability in predictions due to uncertainty in the estimated regression coefficients. By increasing the sample size, we aim to reduce this component of uncertainty and make the model as stable and reliable as possible.

While the primary focus of this article is on epistemic uncertainty, we also included an extension to consider prediction intervals, which incorporate both epistemic and aleatoric uncertainty, the irreducible variability in the outcome that cannot be explained by the model. This total uncertainty sets a lower bound on the achievable precision of predictions, regardless of sample size. In practice, even if model-based uncertainty is minimised, prediction intervals may remain wide as the residual variance is large.

This highlights a fundamental consideration when planning the development or updating of a prediction model: although our method to estimate the sample size required ensures that the model is well-estimated and stable, it cannot overcome limitations imposed by high residual variability. Therefore, researchers must interpret sample size recommendations in the context of the total expected uncertainty in predictions. If the irreducible error is substantial, even a perfectly estimated model may yield predictions that are too imprecise to support clinical decision-making. In such cases, efforts to improve prediction accuracy may need to focus on identifying additional or more informative predictors, rather than simply increasing the sample size, or in some instances, reconsidering whether developing the model is worthwhile at all.

When using the proposed sample size calculations, researchers must consider several practical factors that influence how the methodology is applied in real-world settings. These include defining what constitutes an acceptable level of uncertainty in predictions, selecting appropriate core predictors for the sample size calculation, and understanding how these choices affect the resulting estimates.

A key practical consideration is defining what constitutes an acceptable confidence interval width. This width will vary depending on the clinical context, the outcome being predicted (and the units of measurement), and the consequences of decision-making based on the model. We recommend that researchers engage with clinicians, patients, and other stakeholders to establish context-specific confidence interval widths for acceptable uncertainty.

Another important consideration is the choice of core predictors used in the sample size calculation. If additional predictors are included during model development, particularly those with low prevalence or weak associations, this may increase the uncertainty of predictions. Sensitivity analyses should be conducted to explore the impact of different predictor sets and assumptions about their joint distributions.

One potential barrier to implementation is the need to specify the joint distribution of predictors if researchers do not have access to an existing dataset. However, as demonstrated in our applied example, assuming conditional independence among predictors had minimal impact on the resulting sample size estimates. We also provided pragmatic strategies for approximating joint distributions using published regression coefficients and summary statistics.

When using simulated data, we observed that the maximum required sample size could vary substantially depending on the random seed and the number of simulated observations. The variability was driven by a small number of extreme predictor profiles, which can arise naturally in large, simulated datasets. To mitigate this, we excluded the most extreme 0.001% of simulated individuals (e.g., the top 10 out of 1 million observations), which led to more stable and consistent estimates that closely matched those obtained in the real dataset. This highlights the importance of applying pragmatic thresholds or sensitivity checks when using simulation-based approaches in general for sample size calculations.

In conclusion, we have proposed a novel method for determining the minimum required sample size to develop or update a clinical prediction model with a continuous outcome, with the aim of targeting precise predictions and to inform fairness checks. This approach allows researchers to assess the impact of sample size on prediction uncertainty and to calculate the required sample size to meet a specified level of precision. By supporting both fairness and precision, our method contributes to the development of more reliable and equitable clinical prediction models.